# What One Can Learn From the Cloud Condensation Nuclei (CCN) Size Distributions as Monitored by the BEO Moussala?

V. Kleshtanova[1,2, a)], Ch. Angelov [1, b)], I. Kalapov[1, c)], T. Arsov [d)], G. Guerova [2, e)], V.Tonchev [2, f)]

[1] *Institute of Nuclear Research and Nuclear Energy-BAS, 72 Tsarigradsko shausse blvd., 1784 Sofia, Bulgaria*
[2] *Faculty of Physics, Sofia University, 1164, 5 James Bourchier blvd, 1164 Sofia, Bulgaria*

a) Corresponding author: kleshtanova@gmail.com
b) hangelov@inrne.bas.bg,
c) kalapov@inrne.bas.bg,
d) arsoff@inrne.bas.bg,
e) guerova@phys.uni-sofia.bg,
f) tonchev@phys.uni-sofia.bg

**Abstract.** In this proceeding we report initial studies into the big data set acquired by the Cloud Condensation Nuclei (CCN) counter of the Basic Environmental Observatory (BEO) Moussala over the whole 2016 year at a frequency of 1 Hz. First, we attempt to reveal correlations between the results for CCN number concentrations on the timescale of a whole year (2016) as averaged over 12 month periods with the meteorological parameters for the same period and with the same time step. Then, we "zoom" into these data and repeat the study on the timescale of a month for two months from 2016 – January and July, with a day time step. For the same two months we show the CCN size distributions averaged over day periods. Finally, we arrive at our main result – typical, in terms of maximal and minimal number concentrations, CCN size distributions for chosen hours, one hour for each month of the year, hence 24 distributions in total. These data show a steady pattern of peaks and valleys independent of the concrete number concentration which moves up and down the number concentrations (*y*-axis) without significant shifts along the sizes (*x*-axis).

## I INTRODUCTION

The aerosols are colloid, liquid or solid, particles suspended in a gas phase and in the atmosphere in particular. This is the interaction of the aerosol particles with the clouds that contains at present the greatest uncertainty when estimating the radiative forcing of the Earth atmosphere, see for more details[1], thus preventing from reliable predictions of the climate change[1]. In parallel, the aerosols themselves may have adverse effects on the biosphere as whole and especially on the human health. Their increased concentration leads to albedo change and affects the optical thickness and lifetime of the clouds.

Major property of the aerosols is their ability to facilitate the heterogeneous water condensation (nucleation) thus serving as Cloud Condensation Nuclei (CCN) or, from another context, as *nucleants*. Globally, CCN production has three major sources[2]: "essential" atmospheric nucleation, nucleation around local, non-atmospheric in their nature, sources, both of anthropogenic and natural nature, and "atmospheric processing" of ready but small particles.

Here we report first results of CCN measurements carried at the Basic Environmental Observatory (BEO) Moussala - a modern and complex research facility focused on precision monitoring of the aerospace and terrestrial environment placed at Moussala peak in Rila mountain[3-8] in Bulgaria. The real time measurement of the greenhouse and reactive trace gases that significantly contribute to the thermal budget of the atmosphere thus affecting the climate is one of the key aspects of BEO activities. These as performed by the atmospheric gas analysis system[4]. The observatory is equipped with an automatic weather station comprising basic sensors for air temperature, relative humidity atmospheric pressure, wind speed, etc. Their measurements were validated by comparison with those measured by the NIMH observatory[4] placed also at Moussala peak. BEO is also equipped with instruments for various aerosol measurements – scanning mobility particles sizer, nephelometer, CCN counter.

The CCN counter at BEO Moussala is a commercial CCN Counter[9-17] (CCN-100) produced by Droplet Measurement Technology (DMT).

## II RESULTS

The DMT CCN counter (CCN-100) at BEO Moussala resolves the CCN size distribution in 20 size bins 0.5 μm in width each with data collection frequency of 1 Hz and works in the water supersaturation range of 0.07÷1%. Thus, the major challenge when dealing with such a big data set collected for a year is the data averaging and the concerns of not revealing hidden patterns in the CCN size distributions. Here we start with the information on the CCN total number concentrations as averaged over month periods, FIGURE 1, then "zoom" to see data on the month timescale for two chosen months, January and July, next two figures, averaged over day periods. For these two months we also show the time evolution of the CCN size distribution average over 1-day periods. Finally, we choose for each month of 2016 one hour with minimal and one hour with maximal concentrations of CCN and show the size distribution for this chosen hours, FIGURE 4.

### II.1 CCN concentration during 2016 as compared to the meteorological parameters

To investigate and reveal seasonal trends, the time series of the CCN number concentration, pressure, temperature, humidity and wind velocity are statistically estimated. For this purpose, they are presented monthly using quorums in FIGURE 1. From this type of graphs, we get information about the 25% and 75% percentiles for the smallest (lower quadrant) and the highest (quadrant) numeric value for the median (the value that is located in the middle of the line, i.e. 50% percentile), the mean value of the sample (the arithmetic mean of the data) and the presence of outliers (random data outside the box that are considered as suspicious values). It is clear that there is a seasonal trend. During the warmer half-year (April to September), the CCN concentration is higher than in the colder half-year (November to March). A minimum number of CCN is observed in January, and the maximum is in July, that is why these months are analyzed further in detail. The mean concentrations for these months are 82.74 # / $cm^3$ and 840.46 # / $cm^3$ and the medians are 56.71 # / cm 3 and 812.49 # / cm 3. For comparison, the average annual concentration for 2016 years is 392.87 # / $cm^3$, and the median is 343.13 # / $cm^3$. In January, outliers are closer to the average, and in July they are farther apart, that is, the order of magnitude in January is much smaller than in July. This is due to a more dynamic atmosphere in the summer than in the winter. In each of the astronomical seasons, the mean concentration of CCN is relatively constant.

Concerning the time dependence of the pressure, there are high average values during the summer months (July - September), about 718 hPa and low average values during the winter months (December - February), about 707 hPa. The average pressure for January is 704.87 hPa and for July is 719.31 hPa. The lowest value was recorded on 17.01.2016 - 689.49 hPa, and the highest on 21.06.2016 - 727.17 hPa. The average annual pressure over the period considered is 713 hPa. For comparison in Sofia the normal atmospheric pressure is 948 hPa. The sampling range is larger during the transitional seasons when more cyclones pass over the country.

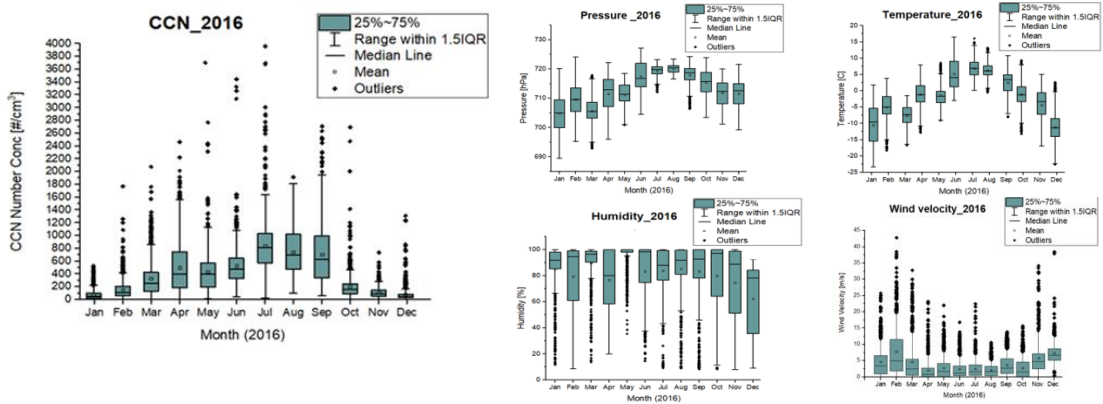

**FIGURE 1** CCN number concentrations (left panel) averaged over the 12 full-month periods of 2016 as correlated with the meteorological parameters pressure, temperature, humidity and wind velocity (right panel)

## II.2 CCN concentration, January and July 2016, vs. meteorological parameters

*II. 2. 1. CCN number concentration (January 2016) vs. the meteorological parameters*

The distribution of the CCN concentration for January shows eight distinct peaks, which are three and more times the average concentration. At six of these peaks the pressure is around and below average, FIGURE 2. Only at two days, on 29.01 and 30.01, the pressure is much higher than the average. At 7-th there is an acute pressure drop associated with cyclone overhead. The other major pressure drop is in 17-th.

The minimum temperature associated with the cold front passage is observed in **FIGURE** 2- evening of 7-th and in the morning of 8-th. In most cases when the temperature is above the mean, the CCN concentration is around the norm or almost zero.

The CCN peaks are observed at humidity higher than the mean, but not the maximal. The humidity throughout the month is close to 100%, observing three deep minimums on the 1-st, 24-th and 31-st when the humidity is below 50 %.

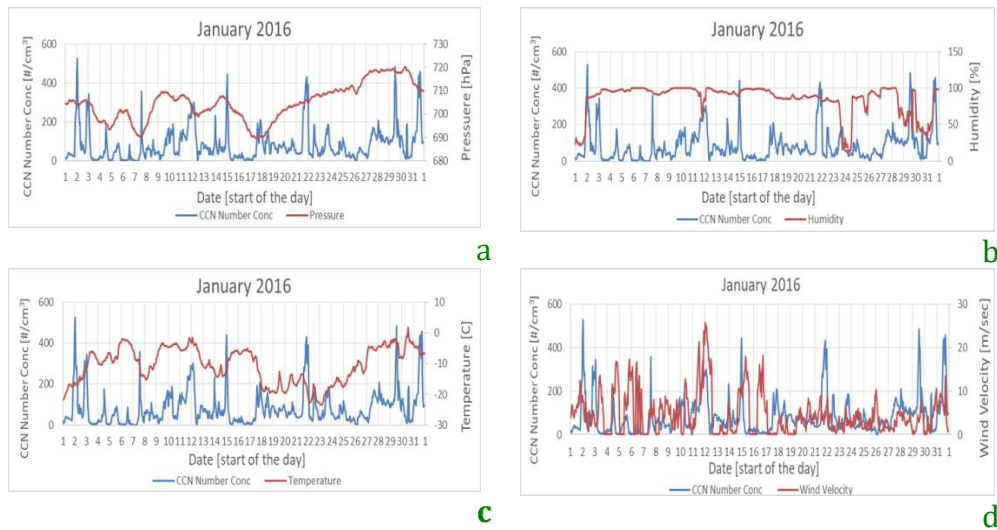

**FIGURE 2** CCN number concentrations for January 2016 (supersaturation 0.58%) as averaged over 31 day periods plotted versus each of the four meteorological parameters monitored: a) pressure; b) humidity; c) temperature and d) wind velocity.

*II. 2. 2. CCN number concentration (July 2016) vs. the meteorological parameters*

The average values of CCN concentrations, pressure, temperature, humidity, wind speed for July are 840.46 # / cm$^3$, 719.31 hPa, 6.82 ° C, 83.75%, 2.54 m / s. All peaks of the CCN concentration are at noon or in evening. The only exception is the peak of the 1-st at 9am. The four largest spikes are on the 5-th, 6-th, 12-th and 13-th of July. The maximum is at 6pm in 13-th - 3960.23 # / cm$^3$. At these peaks a pressure drop is observed. At the highest pressure minimum, at 16-th, the CCN concentration is around the norm. Then the temperature is around the norm - 6,97 ° C and the humidity and the wind are at maximum - 99,9% and 22,47 m / s respectively. Maximum temperature is recorded at 14-th in 12 - 16,03 ° C. Then the temperature begins to fall, passes through a small peak and continues down. At this minimum temperature it is precisely the largest pressure minimum - 712.12 hPa. A slow moving Mediterranean cyclone crosses the country. From July 21 to July 31, there is a very clear twenty-four-hour change of the temperature. During these days, at each peak of the temperature (or around it) a peak of the CCN concentration is observed. The correlation of these parameters over the period considered is 0.49. For comparison, the correlation for the whole month is 0.41. The 24-hour temperature motion leads to the conclusion that there is no dynamic in the atmosphere, but there are local influences. Such influences, for example, are the radiation factors. As in January and July, humidity is very high, during 1/3 of July the humidity is 99.9%. The highest humidity minima are on July 8th and 11th, when it falls to 10.73% and 7.99% respectively.

The wind speed in most days is less than 5 m / s. In 75% of cases it is lower than the average for the month. The wind speed did not affect the concentration of CCN, as it was in January. There are days where the wind speed is minimal (1.25 m / s, at 6 pm at 13 am) and the maximum concentration of CCN is reported. On other days when the wind is maximum (22.47 m / s, at 16.07 at 00 h), the concentration of CCN is also well above the norm - 1470 # / cm$^3$. The corresponding air masses probably come from different places. The first air mass comes from eastern Bosnia and Herzegovina and the second from the Aegean Sea. With this we can explain the big difference between CCN concentrations - when we have sea air mass the concentration of CCN is much smaller than the continental.

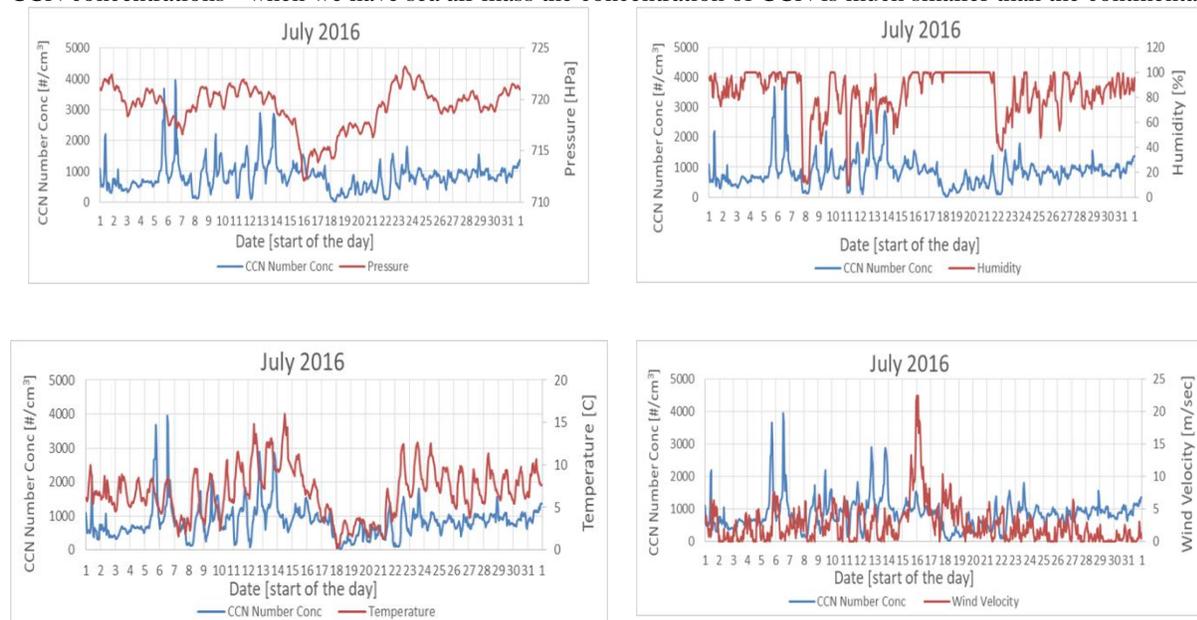

**FIGURE** 3 CCN number concentrations for July 2016 (supersaturation 0.55%) as averaged over 31 day periods plotted versus each of the four meteorological parameters monitored: a) pressure; b) humidity; c) temperature and d) wind velocity.

## II. 3 CCN size distributions

The size distributions of CCN during January and July 2016 are presented in FIGURE 4. It shows that there is almost constant presence of nuclei with sizes between 1.5 and 2.5 microns and between 5 and 6 microns. These dimensions can be defined as "background". From the 15-th to the 20-th bin, i.e. the nuclei larger than 7 microns are almost empty in both months and are not represented.

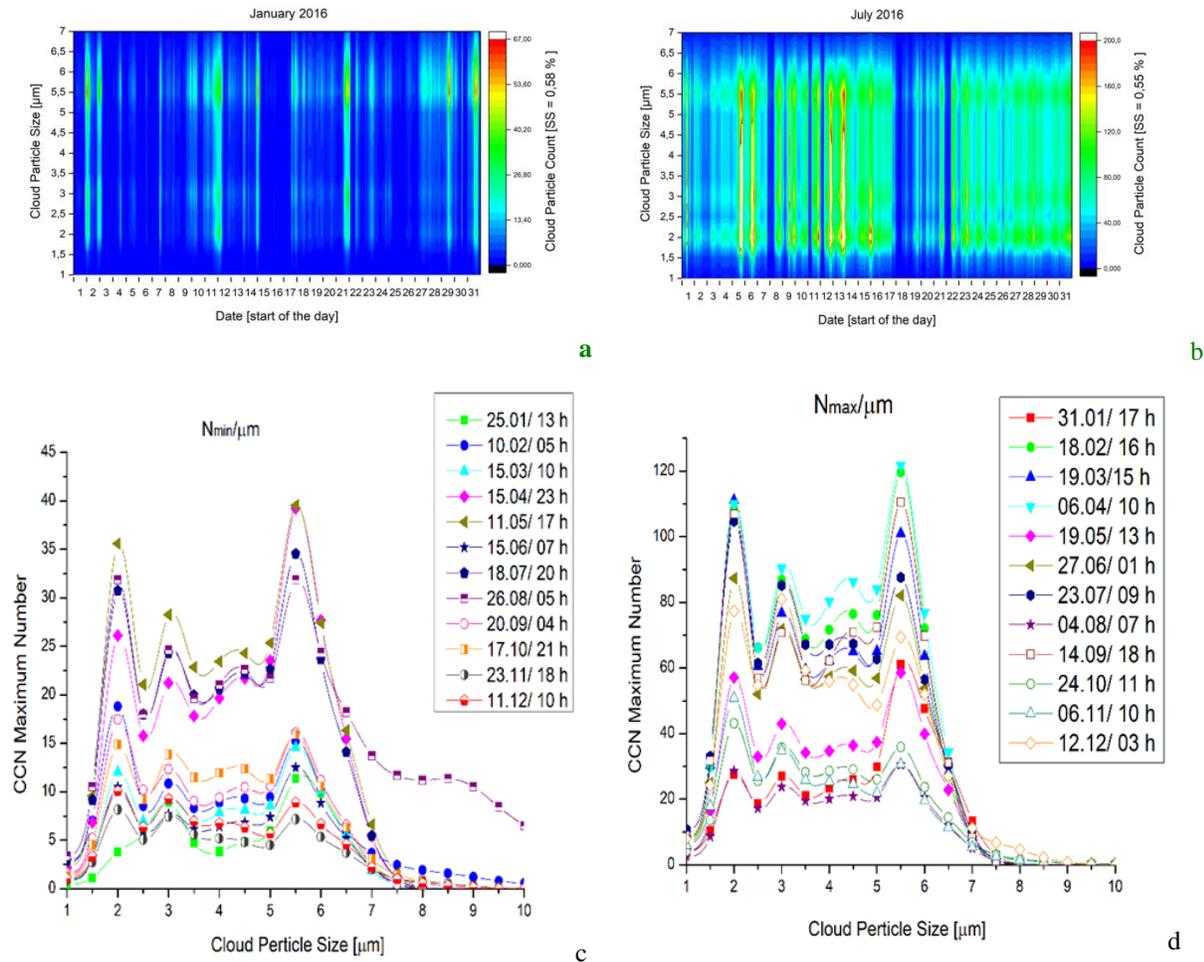

**FIGURE 4** CCN size distributions – over the months of January 2016, (a), and July 2016, (b). In (c) are shown size distributions with minimal number concentrations, one hour per each month of the year, in (d) are shown maximal number concentrations one hour for each of the 12 months.

The figure shows a 24-hour pattern, especially in July. A larger number of nuclei are observed in the afternoon and late hours. In January when the air is cleaner, there are days when many bins are virtually empty. For example, from 14am to 17pm. There are such days in July. In January, a significantly lower number of nuclei (compared to July) was reported - up to 67 units in a single bin. In July, the presence of CCN is significantly higher - up to 573. In the figure on the color palette we can see the distribution of CCN to 200 (red) and more than 200 (white) for the number of nuclei. The latter are individual cases and reported on the 5th, 6th, 12th and 13th. One of these cases is discussed in more detail in the next subsection 3.5. The maximum number of nuclei in January was reported on the 29th at 3 pm - 66 pieces. They fall into the 11th bin and have a size between 5.0 and 5.5 microns respectively. In July, a maximum of the number of nuclei was recorded at 6 pm at 1 pm - 523 counts falling in the 4th bin, ie. with sizes between 1.5 and 2.0 microns. These results confirm the predominant presence of the corresponding bins. This is clearly shown in the figure. It presents separate hours during the year, from different parts of the day. Hours are selected from the monthly size distribution of the nuclei so for $N_{min}$ hours are selected with a minimum number of CCN, and for $N_{max}$ - with a maximum, FIGURE 4, c and d, but more studies are needed on the timescale of seconds in order to elucidate the reasons for the observed behavior.

## CONCLUSIONS

The results from the CCN counter measurements could be linked to a subset of the typical meteorological parameters. There is a steady (invariant) size-distribution pattern that moves up and down the number concentration

($y$-axis) but not along the sizes ($x$-axis). The origin of such invariance is a major challenge for our future work with the data collected from BEO Moussala. Results from Cellular Automata modelling of distributions resulting from heterogeneous nucleation and subsequent growth that is now in progress will be reported elsewhere.

## ACKNOWLEDGMENTS

This work was supported in part by the National Science Fund of Republic of Bulgaria through the project DN 04/1, 13.12.2016:"Study of the combined effect of the natural radioactivity background, the UV radiation, the climate changes and the cosmic rays on model groups of plant and animal organisms in mountain ecosystems" and ACTRIS-2 IA project "Aerosols, Clouds, and Trace gases Research Infrastructure", H2020 (2015-2019). VT is MC member of COST action CM 1402 Crystallize.